%% file: main.tex
\documentclass[sigconf,natbib=true,anonymous=false,nonacm]{acmart}
\usepackage{enumitem}
\usepackage{multirow}
\usepackage{placeins} 
\usepackage{CJK}
\usepackage{float}
\acmConference[
  SIGIR'26\hspace{-0.5mm}      %
]{
  The 49th International ACM SIGIR Conference on Research and Development in Information Retrieval  % 
}{
  \hspace{-0.8mm}July 20-24,2026\hspace{-0.5mm}  % 
}{
  \hspace{-0.8mm}Melbourne, Australia\hspace{-1.1mm}      % 
}

\acmDOI{}  % 录用后填DOI
\acmISBN{} % 录用后填ISBN

\author{
Yifei Chen
}
\affiliation{
  \institution{JD.COM}
  \city{Beijing}
  \country{China}
}
\email{
chenyifei44@jd.com
}
\author{
Zhixing Tian
}
\affiliation{
  \institution{JD.COM}
  \city{Beijing}
  \country{China}
}
\email{
tianzhixing2017@gmail.com
}

\author{
Chenyang Wang
}
\author{
Ziguang Cheng
}
\affiliation{
  \institution{JD.COM}
  \city{Beijing}
  \country{China}
}
\email{
wangchenyang3@jd.com
}
\email{
chengziguang1@jd.com
}

\ccsdesc[500]{Information systems~Information retrieval}
\ccsdesc[300]{Computing methodologies~Natural language processing}

\title{
K-CARE: Knowledge-driven Symmetrical Contextual Anchoring and Analogical Prototype Reasoning for E-commerce Relevance
}  % 
\begin{abstract}
This paper targets e-commerce search relevance. While Large Language Models (LLMs) have demonstrated significant potential in this field, they often encounter performance bottlenecks in persistent "corner cases" within complex industrial scenarios. Existing research primarily focuses on optimizing reasoning trajectories via Reinforcement Learning. However, real-world observations suggest that the primary bottleneck stems from knowledge boundaries, where the absence of domain-specific intelligence in the model’s parametric memory creates a contextual void. This void persists when interpreting idiosyncratic queries or niche products and cannot be resolved solely through reasoning-path optimization.

To bridge this gap, we propose K-CARE, a framework that extends the model’s cognitive reach by grounding reasoning in external knowledge. K-CARE comprises two synergistic components: (1) Symmetrical Contextual Anchoring (SCA), which fills the contextual void by anchoring queries and products with behavior-derived implicit knowledge; and (2) Analogical Prototype Reasoning (APR), which leverages expert-curated prototypical knowledge to calibrate decision boundaries through in-context analogy. Extensive offline evaluations and online A/B tests on a leading e-commerce platform demonstrate that K-CARE significantly outperforms state-of-the-art baselines, delivering substantial commercial impact by resolving knowledge-intensive relevance challenges.

\end{abstract}
\keywords{Text Classification, E-commerce Retrieval, Search Relevance}  % 关键词

\begin{document}
\maketitle  % 生成标题/作者/摘要

% 引用你的子文件（chapter/intro.tex）
\input{chapter/intro_v3}
% \input{chapter/framework}
\input{chapter/related_work}

\input{chapter/methodology_v3}
\input{chapter/experiments_v2}

\input{chapter/Conclusion_v2}
% 其他子文件引用（按需添加）
% \input{chapter/method.tex}
% \input{chapter/experiment.tex}

% --------------------------
% 6. 参考文献（SIGIR要求ACM格式）
% --------------------------
\bibliographystyle{ACM-Reference-Format}  % SIGIR指定样式
\bibliography{nime-references}  % 你的bib文件（无需改名，兼容）

\end{document}

%% file: chapter/intro_v3.tex
\section{Introduction}
Search serves as the primary gateway in e-commerce, requiring the relevance module to accurately match concise query strings against massive product inventories. The precision of relevance is the critical determinant of user experience. 
To handle this matching task, the field has increasingly transitioned from BERT-style models~\cite{DBLP:conf/naacl/BertDevlinCLT19,DBLP:journals/corr/abs-1907-11692-roberta, DBLP:conf/iclr/HeLGC21-deberta,DBLP:conf/iclr/LanCGGSS20-albert,DBLP:journals/corr/alibaba-robust-interaction-based-bert,DBLP:conf/kdd/reprBERT, DBLP:conf/nips/LewisPPPKGKLYR020-rag-grandfather} to LLMs~\cite{DBLP:journals/corr/serm-bytedance-self-evolving-relevance-llm,DBLP:journals/corr/meituan-reasoning-llm-bert,alibaba-ad-think-broad-acting-fast,DBLP:journals/corr/pinterest-relevance-llm,DBLP:journals/corr/microsoft-llm-search-relevance,DBLP:conf/kdd/tencent-search-relevance-llm,DBLP:conf/www/lref-jd, DBLP:journals/corr/abs-2508-12365-taosr1, DBLP:journals/corr/abs-2510-07972-taosr-she, DBLP:journals/corr/abs-2510-08048-taosr-agrl,DBLP:journals/corr/taobao-lore,DBLP:journals/corr/walmart-llm-search-relevance}. The problem is thus framed as a relevance reasoning chain, typically involving query and product understanding and matching rubric application, and those LLM-based approaches have demonstrated notable efficacy. 

Nevertheless, for modeling relevance in real-world industrial scenarios involving vast product inventories and highly diverse user expressions, current LLM-based methods still encounter significant performance bottlenecks in persistent corner cases. To mitigate these performance bottlenecks, industry practitioners have turned to adopting Reinforcement Learning (RL), such as DPO~\cite{DBLP:conf/nips/RafailovSMMEF23-DPO}, KTO~\cite{DBLP:journals/corr/kto-paper} and GRPO~\cite{DBLP:journals/corr/abs-2402-03300-deepseek-math}, to optimize the relevance reasoning chain by eliciting superior reasoning trajectories~\cite{DBLP:conf/www/lref-jd,DBLP:journals/corr/abs-2508-12365-taosr1,DBLP:journals/corr/abs-2510-07972-taosr-she,DBLP:journals/corr/abs-2510-08048-taosr-agrl,DBLP:journals/corr/taobao-lore,alibaba-ad-think-broad-acting-fast,DBLP:journals/corr/meituan-reasoning-llm-bert}. 
However, observations from real-world scenarios suggest that the primary bottleneck is not the pathway of reasoning, but the boundary of knowledge. Specifically, LLMs often encounter `blind spots' where the requisite context for interpreting idiosyncratic queries or niche products is absent from their parametric memory. Our error analysis indicates that over 60\% of persistent failures stem from such contextual voids, which cannot be resolved by simply optimizing reasoning trajectories 
% (e.g., via DPO, KTO or GRPO)
without grounding them in domain-specific intelligence.
% ~\cite{DBLP:journals/corr/abs-2504-13837-judge-rl}.

%over 60\% 根据badcase，粗算比例为62%
\begin{figure*}[t] % [t] 表示将图片放置在页面的顶部 (top)
    \centering
    % \hspace{100cm}
\includegraphics[width=1.08\textwidth]{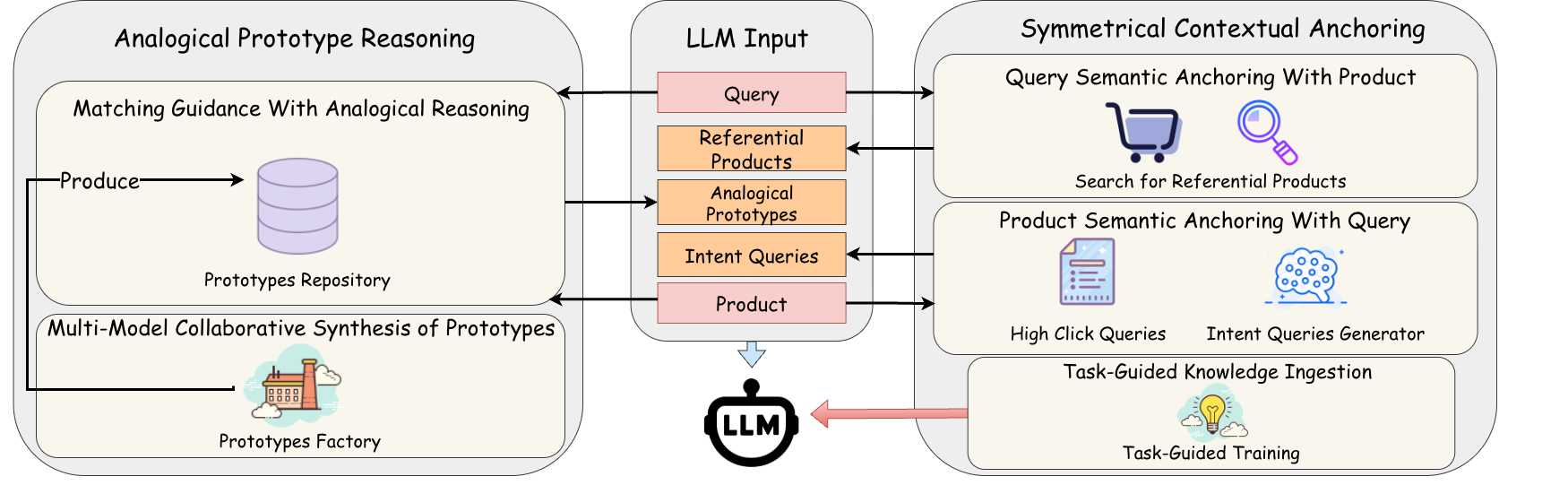} 
    % \hspace{0cm}
    % \caption{framework of K-CARE}
    \caption{K-CARE: (1) Symmetrical Contextual Anchoring (SCA) grounds reasoning in behavior-derived implicit knowledge through bidirectional anchors. (2) Analogical Prototype Reasoning (APR) leverages expert-curated prototypical knowledge to calibrate decision boundaries via in-context analogy}
    \label{fig:kcare_framework}
\end{figure*}

To bridge this gap, we propose K-CARE (\textbf{K}nowledge-driven symmetrical  \textbf{C}ontextual Anchoring and \textbf{A}nalogical Prototype \textbf{RE}asoning), 
a framework designed to extend the model's cognitive reach by incorporating \textbf{behavior-derived implicit knowledge} that captures latent semantic associations and \textbf{expert-curated prototypical knowledge} that aligns with nuanced matching rubrics. 
K-CARE operationalizes this knowledge-driven approach through two synergistic components:
(1) \textbf{Symmetrical Contextual Anchoring (SCA)}: to resolve semantic interpretability, we implement a symmetrical grounding mechanism where queries are interpreted through their associated products and products are refined via their potential search intents. Recognizing that such behavioral knowledge is inherently sparse or noisy in industrial logs, we resolve this by leveraging search system ranking signals to collect referential products for queries and employing a predictive product-to-query model to generate intent queries for products. This dual-path enrichment provides high-coverage, knowledge-dense semantic anchors for both entities. Furthermore, SCA incorporates Task-Guided Knowledge Ingestion (TGKI) to ensure the LLM effectively leverages this enriched context through specialized training objectives.
(2) \textbf{Analogical Prototype Reasoning (APR)}: to facilitate reasoning calibration and align with matching rubrics, we construct a repository of expert-curated and informative exemplars paired with explicit rationales. 
By injecting these high-quality prototypes into the reasoning context
as interpretative guides, K-CARE enables the LLM to perform analogical reasoning, effectively calibrating its decision boundaries in ambiguous scenarios. 

We evaluate K-CARE through extensive offline experiments and online A/B tests on a leading e-commerce platform JD.com. The results demonstrate that K-CARE significantly outperforms state-of-the-art baselines, resolving complex, knowledge-intensive relevance challenges with substantial commercial impact. The primary contributions of this work are summarized as follows: 

\begin{itemize}
    \item \textbf{The K-CARE Framework}: We propose a novel, knowledge-driven framework that extends the cognitive reach of LLMs beyond parametric memory, specifically tailored for complex industrial e-commerce relevance tasks.
    \item \textbf{Symmetrical Contextual Anchoring}: We introduce a symmetrical grounding mechanism that leverages behavior-derived
    implicit knowledge to enhance semantic interpretability for both queries and products while mitigating the impact of feedback sparsity.
    \item \textbf{Analogical Prototype Reasoning}: 
    We develop an analogical reasoning approach that leverages expert-curated prototypical knowledge to calibrate decision boundaries through in-context analogy, facilitating robust and nuanced relevance judgments in complex scenarios.
\end{itemize}

%  北极星  matching criteria 和matching rubrics似乎表达的意思是上有一定的区别 需要确认 是否需要对齐 

%% file: chapter/related_work.tex
\section{Related Work}
Traditional methods like BM25 \cite{DBLP:conf/trec/bm25} and TF-IDF \cite{DBLP:journals/jd/TfIdfJones,DBLP:journals/cacm/SMARTSaltonL65} can capture text matching between search queries and documents information, but they fail to model semantic similarity. 
With the advancement of deep learning techniques, DSSM \cite{DBLP:conf/cikm/DSSM} is proposed for handling semantic level matching. However, as DSSM is a variation of the Bag-of-Words model based on word hashing, it maps input text into static vector representations which could lead to the confusion of polysemous words.
While DSSM is constrained by static word hashing, the advent of the attention mechanism has enabled models to capture intricate contextual dependencies.
Transformers \cite{DBLP:conf/nips/attention_is_all_you_need} based encoder architectures have been employed to model contextualized semantic dependencies in search scenarios. In e-commerce search scenarios, pre-trained models such as BERT \cite{DBLP:conf/naacl/BertDevlinCLT19} and its variants~\cite{DBLP:conf/naacl/BertDevlinCLT19,DBLP:journals/corr/abs-1907-11692-roberta, DBLP:conf/iclr/HeLGC21-deberta,DBLP:conf/iclr/LanCGGSS20-albert, DBLP:journals/corr/alibaba-robust-interaction-based-bert, DBLP:conf/kdd/reprBERT}  have been widely adopted for relevance tasks. 
% For instance, ReprBERT \cite{DBLP:conf/kdd/reprBERT} employs knowledge distillation to transfer semantic insights from a cross-encoder teacher into an efficient dual-tower student, effectively balancing high precision with low-latency search requirements. In contrast, ei-SRC \cite{DBLP:journals/corr/alibaba-robust-interaction-based-bert} adopts a lightweight 3-layer interaction-based architecture, utilizing a dynamic-length representation scheme and contrastive adversarial training to maintain.
Specifically, ReprBERT \cite{DBLP:conf/kdd/reprBERT} leverages knowledge distillation to transfer complex semantic insights from a cross-encoder teacher into an efficient dual-tower student, effectively balancing high precision with low-latency requirements. Meanwhile, ei-SRC \cite{DBLP:journals/corr/alibaba-robust-interaction-based-bert} adopts a lightweight 3-layer interaction-based architecture, incorporating a dynamic-length representation scheme and contrastive adversarial training to preserve deep semantic interactions while significantly reducing inference overhead.

In recent years, LLMs have attracted increasing attention for e-commerce search relevance due to their expanded parameter scale and enhanced ability to leverage contextual information
\cite{DBLP:conf/iclr/janpan_anal_reasoner,DBLP:conf/nips/nips_context_learning}. Compared with BERT-style encoders, LLMs generally offer larger model capacity and stronger contextual semantic modeling, which can benefit intent understanding and query–document matching, and has led to growing interest in integrating LLMs into relevance modeling and ranking. 
Existing studies mainly follow two directions: improving supervised fine-tuning (SFT) and training data construction to better align models with relevance criteria and preferences~\cite{DBLP:conf/www/lref-jd}, and exploring reinforcement learning (RL) paradigms to enhance the stability and consistency of relevance reasoning chains~\cite{DBLP:journals/corr/abs-2508-12365-taosr1,DBLP:journals/corr/abs-2510-07972-taosr-she,DBLP:journals/corr/abs-2510-08048-taosr-agrl}.
On the SFT centric line of work, LREF~\cite{DBLP:conf/www/lref-jd} explicitly targets three deployment bottlenecks—noisy labels, uncontrollable reasoning, and optimistic bias on borderline cases by combining selection-driven SFT with Multi-CoT tuning and DPO-based de-biasing, thereby improving both data reliability and decision calibration.
On the SFT centric line of work, LREF~\cite{DBLP:conf/www/lref-jd} explicitly targets three deployment bottlenecks, namely noisy labels, uncontrollable reasoning, and optimistic bias on borderline cases, by combining selection driven SFT with Multi-CoT tuning and de-biasing via DPO, thereby improving both data reliability and decision calibration.
Beyond data and preference alignment, recent RL-based studies focus on making the reasoning process itself more robust under real-world search constraints. TaoSR1~\cite{DBLP:journals/corr/abs-2508-12365-taosr1} takes an early step by reducing CoT error accumulation and discriminative hallucination, integrating post-CoT processing with DPO and difficulty-aware GRPO to support efficient online deployment. Building on this, SHE~\cite{DBLP:journals/corr/abs-2510-07972-taosr-she} further improves long-tail generalization by introducing SRPO with hybrid step-level rewards, complemented by diversified filtering and curriculum learning to mitigate sparse and noisy feedback. Meanwhile, TaoSR-AGRL~\cite{DBLP:journals/corr/abs-2510-08048-taosr-agrl} emphasizes rule-compliant multi-step inference under complex business constraints, and alleviates sparse terminal rewards and slow convergence via rule-aware reward shaping and adaptive guided replay.
% From a broader perspective, relevance judgment typically relies on two capabilities: (i) adhering to matching criteria with controllable reasoning, and (ii) leveraging sufficient domain knowledge to interpret long-tail products and idiosyncratic user expressions. In comparison, SFT and RL more directly improve the former, while purely parametric memory often falls short when facing rapidly evolving knowledge and long-tail coverage issues in e-commerce. 
While these approaches progressively enhance reasoning quality and alignment with matching rubrics, they share a common limitation: all optimization occurs within the model's parametric knowledge, leaving the knowledge boundary unexpanded. When the model lacks the domain-specific context to interpret an idiosyncratic query or a niche product, reasoning-path optimization alone cannot compensate for this void.
To mitigate the staleness and incompleteness of parametric knowledge, a common line of work augments model judgments with external knowledge. In this regard, retrieval-augmented generation~\cite{DBLP:conf/nips/LewisPPPKGKLYR020-rag-grandfather,DBLP:journals/corr/abs-2312-10997-rag-survey, DBLP:conf/icml/improve_llm_from_retrieve, DBLP:conf/emnlp/llm_with_citation} provides a representative paradigm to solve the contextual avoid through knowledge injection. However, effectively instantiating RAG in a specific domain requires identifying what knowledge to retrieve and how to enable the model to leverage context.
In the context of e-commerce relevance, our error analysis reveals two distinct types of knowledge gaps: semantic voids, where the model lacks the context to interpret idiosyncratic queries or niche products, and ambiguous decision boundaries, where the model struggles with nuanced matching rubrics despite adequate semantic understanding.
These findings motivate the design of K-CARE as a domain-specific realization of the RAG paradigm: SCA retrieves behavior-derived implicit knowledge to fill semantic voids through bidirectional anchoring, while APR retrieves expert-curated prototypical knowledge with explicit rationales to demonstrate matching rubrics through analogical reasoning.

%% file: chapter/methodology_v3.tex
\section{Methodology}

We formally define the search relevance task as a supervised classification problem. Given a user-issued query $Q$ and a candidate product $P$, the goal is to learn a mapping function $f: (Q, P) \rightarrow Y$ that predicts the relevance degree $y \in Y$. The label space $Y$ consists of three distinct categories, specifically: Perfect, Passable and Bad. The objective of the model $f$ is to estimate the conditional probability $P(y|Q, P)$ to accurately characterize the relevance between the query $Q$ and the product $P$.

% \vspace{-2.6mm}

\subsection{Overview}
For e-commerce search relevance, we propose K-CARE, a knowledge-driven reasoning framework to fill the contextual voids arising from the absence of critical domain knowledge within an LLM's parametric memory.
As illustrated in Fig.~\ref{fig:kcare_framework}, K-CARE consists of two synergistic components.

% 北极星 intro部分是 domain-specific implicit knowledge 和 expert-curated prototypical knowledge 加粗并列 但是这里变成了  behavior-derived 变成了一个修饰前缀 需要确认有无问题

First, we introduce \textbf{Symmetrical Contextual Anchoring (SCA)}, 
which leverages behavior-derived implicit knowledge to enhance semantic context for both queries and products.
% which enriches the query and product inputs with collective search intelligence, producing a knowledge-dense context to facilitate deeper semantic comprehension.
Crucially, SCA incorporates Task-Guided Knowledge Ingestion (TGKI), using specialized training tasks to ensure the LLM effectively absorb this enriched context.
Building upon this strengthened comprehension, we further propose \textbf{Analogical Prototype Reasoning (APR)}, which leverages expert-curated prototypical knowledge to calibrate decision boundaries through in-context analogy.
% By injecting expert-curated prototypes into the reasoning context, APR aligns the model's inference with expert matching rubrics and effectively calibrates its decision boundaries in ambiguous scenarios.

% \vspace{-2.6mm}

\subsection{Symmetrical Contextual Anchoring}
SCA consists of two primary stages: knowledge acquisition and knowledge ingestion. For acquisition, we develop Query Semantic Anchoring with Product (QSAP) and Product Semantic Anchoring with Query (PSAQ) 
% to establish bidirectional semantic links. 
to enrich the context with behavior-derived implicit knowledge through bidirectional semantic links.
To further ensure the LLM effectively leverages this enriched context, we introduce Task-Guided Knowledge Ingestion (TGKI).

% \vspace{+1.5mm}
\textbf{Query Semantic Anchoring With Product (QSAP)} enhances the LLM’s comprehension of the query with referential products.
% The latent intent of a concise or ambiguous query can be explicitly decoded through the attributes and categories of its corresponding product set.
While LLMs are proficient at interpreting literal semantics, they often lack the domain-specific conceptual knowledge required to resolve long-tail or idiosyncratic terms in e-commerce search. For instance, recognizing that a colloquial nickname refers to a specific brand, or that a descriptive phrase corresponds to a particular product category, remains beyond the model's reach. The search system, however, inherently embeds rich behavioral knowledge that exceeds the model's cognitive reach, as it continuously aggregates and reflects real user engagement.
A straightforward strategy for this anchoring is to leverage historical click-through data, using top-clicked products as auxiliary context. 
However, this feedback-dependent approach suffers from a "cold-start" limitation where click logs for tail or new queries are inherently sparse.
To overcome this, we utilize the e-commerce search system’s ranking mechanism as a "silver standard," retrieving top-ranked products from the entire inventory to serve as robust contextual anchors. In practice, each query is submitted to the live search system, and the returned ranking output constitutes the retrieved anchor set.
Although the ranking output inevitably contains some noise, LLMs possess strong contextual understanding and noise-tolerance capabilities, enabling them to extract effective semantic signals from noisy anchors without being misled. This complementarity between the coverage advantage of the silver standard and the robustness of LLMs ensures that even queries with zero click history can obtain reliable semantic anchoring.

% 北极星 这里的 relevant products 是否需要改成 referential products

% \vspace{+1.5mm}

\textbf{Product Semantic Anchoring With Query (PSAQ)} 
symmetrically improves the LLM’s comprehension of the product with intent queries. 
% Just as queries are clarified by products, a product’s precise utility is best defined by the specific queries that trigger user engagement.
Just as queries are clarified by products, a product's implicit functional semantics may not be fully captured by its attribute description, which is authored from the seller's perspective and inherently biased toward marketing-oriented expressions. The customer's perspective, by contrast, reveals the product's functional identity as defined by actual demand rather than promotional claims.
While a straightforward strategy involves injecting highly-clicked queries as supplementary signals, this similarly faces sparsity issues for niche or long-tail items.
To resolve this, we train a product-to-query (P2Q) prediction model that forecasts potential high-intent queries when historical data is insufficient. The P2Q model is a generative model trained on historical click logs, producing up to approximately five to six intent queries per product.
This integration of historical feedback and predictive modeling overcomes the inherent sparsity of interaction logs, ensuring that even niche or long-tail products are equipped with informative semantic anchors. 

% 北极星 这里的 relevant queries 是否需要改成 intent queries

\begin{table*}[!htbp]
  \centering
\caption{
        Evaluation Results 
  }
  \setlength{\tabcolsep}{1.8mm}{
      \begin{tabular}{c|ccc|ccc}
                \toprule
                \multirow{2}{*}{\textbf{Models}}  
                % & \multirow{2}{*}{\textbf{Components}}
                & \multicolumn{3}{c|}{\textbf{Three-Level Classification}}
                % & \multicolumn{3}{c|}{\textbf{ Relevant Classification}} 
                & \multicolumn{3}{c}{\textbf{ Irrelevant Classification}} 
                \\
                &\textbf{Macro F1} &\textbf{Weighted F1} &\textbf{Accuracy}
                % &\textbf{Precision} &\textbf{Recall} &\textbf{F1}
                &\textbf{Precision} &\textbf{Recall} &\textbf{F1}
                \\
                \midrule 
                LLM Base    
                % & -
                & 52.94  & 60.29  & 59.73 
                % & 64.51 & \textbf{97.28} &77.58 
                &\textbf{92.42} &39.72 &55.56
                \\
                LLM SFT     
                % & -
                & 80.97  & 86.82  & 86.95 
                % & 91.70 & 91.37 &91.54 
                & 89.91   &90.30 &90.11 
                \\
                LLM GRPO    
                % & - 
                & 81.18  & 87.04  & 87.17 
                % & 92.47 & 91.02 &91.74 
                &89.66 &91.30 &90.48
                \\
                \midrule 
                K-CARE w/ SCA (QSAP)     
                & 82.00 &  87.53 & 87.61  
                % & 92.14 & 92.09& 92.11 
                &90.70 &90.78 &90.74
                \\
                K-CARE w/ SCA (QSAP+PSAQ)          
                & 82.16 &  87.66 & 87.73  
                % & 92.34 & 91.96& 92.15 
                &90.61 &91.07 &90.84
                \\
                K-CARE w/ SCA (QSAP+PSAQ+TGKI)    
                & 82.39 &  87.85 & 87.94  
                % & 92.68 & 91.80& 92.24 
                &90.49 &91.50 &90.99
                \\
                \midrule 
                K-CARE w/ SCA + APR (Proposed)
                & \textbf{82.58} &  \textbf{87.99} & \textbf{88.05}  
                % & \textbf{92.71} & 91.98& \textbf{92.34} 
                &90.68 &\textbf{91.51} &\textbf{91.09}
                \\
                \bottomrule
        \end{tabular}
    }
      
    \label{tab:experiment_overall}
\end{table*}

%\subsubsection{reasoning sub-trajectory alignment}
%\hfill \\\
% \vspace{+1.5mm}
\textbf{Task-Guided Knowledge Ingestion (TGKI)}: 
While QSAP and PSAQ provide rich contextual anchors, this information is often implicit and indirect (e.g., product lists rather than explicit definitions). Consequently, the LLM faces a significant cognitive burden in inferring underlying semantics.
We introduce Task-Guided Knowledge Ingestion (TGKI). The core of TGKI is a set of auxiliary training tasks $\mathcal{T} = \{t_{q\_und}, t_{q\_pre}, t_{p\_und}, t_{p\_pre}\}$, specifically designed to explicitly guide the model in deciphering and internalizing the latent semantic links within the provided context:
\begin{itemize}
    \item \textbf{Query Understanding ($t_{q\_und}$):} Interpreting query intent via top-ranked products.
    \item \textbf{Query Prediction ($t_{q\_pre}$):} Reconstructing the query from top-ranked product context.
    \item \textbf{Product Understanding ($t_{p\_und}$):} Synthesizing product profiles from associated queries.
    \item \textbf{Product Prediction ($t_{p\_pre}$):} Predicting intrinsic product properties based on user intent.
\end{itemize}
Table~\ref{tab:tgki_tasks} details the prompt template for each task.
For each task $t \in \mathcal{T}$, we construct a synthetic dataset $\mathcal{D}_t = \{(x_i, y_i)\}$, where reasoning trajectories $y_i$ are synthesized by a high-capacity teacher model (e.g., GPT-5.2). 
These tasks are designed along two complementary dimensions. The understanding tasks ($t_{q\_und}$, $t_{p\_und}$) train the model to internalize the interaction patterns with implicit behavioral knowledge, learning how to effectively extract and utilize the information conveyed by contextual anchors. Building upon this foundation, the prediction tasks ($t_{q\_pre}$, $t_{p\_pre}$) train the model to comprehend the injected contextual anchors, enabling it to leverage this information for relevance judgment. 
This process familiarizes the model with the interaction patterns of the information provided by QSAP and PSAQ, thus warming up the model's capability to leverage indirect behavioral knowledge. 
To prevent gradient interference between these diverse ingestion objectives and the core relevance task, we adopt a decoupled training strategy. We first optimize the model to establish a solid perceptual foundation by training on the four TGKI tasks jointly, followed by fine-tuning on the core relevance task, thereby ensuring that the ingested knowledge is effectively translated into precise relevance judgments.

\begin{table}[htbp]
  \centering
  \small
  \caption{Prompt templates for TGKI training objectives.}
  \label{tab:tgki_tasks}
  \begin{tabular}{c|l}
    \toprule
    \textbf{Task} & \textbf{Prompt Template} \\
    \midrule
    $t_{q\_und}$ & \textit{Given query and its top-ranked} \\
                  & \textit{products, interpret the query intent.} \\
    \midrule
    $t_{q\_pre}$ & \textit{Given a top-ranked product, predict potential} \\
                  & \textit{search queries.} \\
    \midrule
    $t_{p\_und}$ & \textit{Given a product and its clicked queries,} \\
                  & \textit{synthesize product profile.} \\
    \midrule
    $t_{p\_pre}$ & \textit{Given clicked queries only, predict intrinsic} \\
                  & \textit{product properties.} \\
    \bottomrule
  \end{tabular}
\end{table}
 
% \vspace{-2.5mm}
% \vspace{+1.5mm}

\subsection{Analogical Prototype Reasoning}
While SCA enriches the unilateral comprehension of queries and products, it primarily focuses on entity-level semantics rather than the bilateral matching rubrics. In complex or long-tail scenarios, the relevance judgment often hinges on nuanced matching rubrics that define why a specific product satisfies (or fails) a particular query. Such intricate decision-making requires more than just semantic understanding. To address this, we propose Analogical Prototype Reasoning (APR), which injects expert-curated prototypical cases and their rationales into the reasoning context, enabling the LLM to calibrate its decision boundaries through in-context analogy.

% 北极星 repository 是否需要把图中的dataset 改成 repository 目前看 repository是主流表达
% \vspace{+1.5mm}

\textbf{Multi-Model Collaborative Synthesis of Prototypes}:
the construction of a high-quality prototype repository is essential for effective analogical reasoning. To this end, we introduce a Multi-Model Collaborative Synthesis pipeline designed to automatically generate reasoning quadruplets $\mathcal{S} = \langle q, p, l, r \rangle$. Each quadruplet consists of a challenging query-product pair $(q, p)$, a reliable relevance label $l$, and a logically coherent rationale $r$.
The synthesis process is operationalized through the following four-stage pipeline:
\begin{itemize}

% \textbf{Stage I: Hard Case Mining}:
\item \textbf{Hard Case Mining}: 
to ensure the repository focuses on high-value prototypes for ambiguous scenarios, we employ a sub-optimal model $M_{sub}$ to identify challenging instances where the model's prediction $l_{sub}$ conflicts with the ground-truth human label $l$. 
\begin{equation}
    D_{hard} = \{ (q, p, l) \in D_{human\_labeled} \mid M_{sub}(q, p) \neq l \}
\end{equation}
$D_{hard}$ are then passed to the next stage, which ensures the repository focuses on challenging and informative boundary cases.

% \textbf{Stage II: Multi-Model Proposal}:
% \vspace{+1.5mm}

\item \textbf{Multi-Model Proposal:} 
to cross-verify human-annotated labels and provide multi-faceted justifications, we employ an ensemble of heterogeneous models $\mathcal{M} = \{M_1, \dots, M_n\}$ as proposers. Specifically, each model $M_i$ generates a candidate proposal $P_i$, consisting of a judgment and a supporting rationale:
\begin{equation}
    P_i = (judge_i, r_i) = M_i(q, p, l), \quad (q, p, l) \in D_{hard}
\end{equation}
where $judge_i \in \{0, 1\}$ indicates the model's assessment of whether the human-annotated label $l$ is correct and $r_i$ is the supporting rationale.

% \vspace{+1.5mm}

\item \textbf{Cross-Proposal Arbitration}: to resolve potential conflicts among heterogeneous proposals and synthesize a unified expert-level reference, we employ a high-capacity model $M_{arb}$ to adjudicate the collective proposals generated in the previous stage:
\begin{equation}
    (judge^*, r^*) = M_{arb}(q, p, l, \mathcal{P}), \quad \mathcal{P} = \{P_1, \dots, P_n\}
\end{equation}
Only instances passing this check ($judge^* = 1$) are compiled into a candidate consensus set $D_{cons} = \{ (q, p, l, r^*) \mid judge^* = 1 \}$

% \vspace{+1.5mm}

\item \textbf{Linguistic Quality Assurance}: to ensure the synthesized rationales are articulate and logically sound, a quality-focused model $M_{qa}$ performs a linguistic audit on each $r^*$ in $D_{cons}$, yielding the final rationale $r_{final}$:
\begin{equation}
    r_{final} = M_{qa}(r^*), \quad \text{s.t. } \text{Logic}(r_{final}) \text{ is coherent.}
\end{equation}
The final prototype repository is thus defined as the set of successfully audited quadruplets:
\begin{equation}
    \mathcal{D}_{proto} = \{ \langle q, p, l, r_{final} \rangle \mid (q, p, l, r^*) \in D_{cons} \}
\end{equation}

\end{itemize}

% \vspace{+1.5mm}

\textbf{Matching Guidance with Analogical Reasoning}: during the inference phase, APR provides the LLM with explicit matching guidance by situating ambiguous cases within the context of these retrieved authoritative prototypes. Specifically, we employ an embedding model to convert the concatenated information of query and product into dense vector representations. A similarity search is then performed to retrieve the most analogous prototype from $\mathcal{D}_{proto}$
 for each target pair. By framing the current task as an analogous counterpart to the retrieved exemplar, APR enables the LLM to map the expert's reasoning trajectory onto the target case, effectively calibrating its decision boundaries in complex relevance scenarios.

 % 北极星 authoritative prototypes 和  analogous prototype 形容词是否需要统一 突然想了一下 感觉不用 authoritative说明 prototypes可用  analogous形容的是 它的推理模式

%% file: chapter/experiments_v2.tex
\section{Experiments}
\subsection{Datasets and Metrics}

\textbf{Datasets: }The test datasets comprise over 120,000 query-product pairs sampled from real-world search logs of JD, covering diverse e-commerce scenarios.
Each instance in the test datasets is refined through multi-turn human verification into three distinct relevance levels: \textit{Perfect}, \textit{Passable}, and \textit{Bad}. 
The distribution of the test datasets is shown in Table \ref{tab:dataset_stats}. 
Notably, the data exhibits a bimodal distribution: Bad and Perfect cases together account for over 88\% of the total, while Passable cases constitute only 11.83\%.  The high proportion of Bad cases (46.02\%) also underscores the prevalence of irrelevant exposures in industrial search scenarios, which motivates our focus on improving Bad case detection as reported in the Irrelevant Classification metrics.

\textbf{Metrics:}
We evaluate the resulting multi-class classification task from two perspectives: (1) \textbf{Three-Level Classification}: we employ Macro F1, Weighted F1, and Accuracy to assess multi-level product tiering performance across all categories. 
A better tiering capability indicates that the model can reliably distinguish fine-grained relevance intensities, which is crucial for ranking because it helps prioritize truly high-relevance products while reducing borderline misjudgments between Passable and Bad.
(2) \textbf{Irrelevant Classification}: we measure the Precision, Recall, and F1-score of the \textit{Bad} category to evaluate the model's capability in filtering irrelevant results and minimizing noisy exposures. We report the \textit{Bad} category metrics to explicitly measure, in an offline manner, the model’s ability to identify bad cases, which is critical for reducing the exposure of irrelevant items in top-ranked positions.

\begin{table}
  \caption{Distribution of Test Datasets}
  \label{tab:dataset_stats}
  \begin{tabular}{ccl}
    \toprule
    Relevance Level & Number&Proportion\\
    \midrule
    Bad & 57,091 & 46.02\%  \\
    Passable & 14,682  & 11.83\%   \\
    Perfect & 52,283 & 42.14\%  \\
    \midrule
    \textbf{Total} & \textbf{124,056} & \textbf{100.00\%} \\
  \bottomrule
\end{tabular}
\end{table}

% 北极星  ranking 感觉是不是还是只有分层就好 这也不排序啊 包括online study 写的也是 tiering

\vspace{-0.5mm}

\subsection{Baselines}
We compare K-CARE with several representative baselines initialized from Qwen3-8B~\cite{DBLP:journals/corr/qwen3-tec-report}:

\noindent\textbf{LLM Base:} The vanilla Qwen3-8B model is evaluated directly to establish its zero-shot performance on relevance task.

\noindent\textbf{LLM SFT:} We SFT LLM Base on our high-quality relevance datasets to adapt its parametric memory to the task.

\noindent\textbf{LLM GRPO:} Building on LLM SFT, we apply GRPO to further optimize reasoning trajectories within the parameters.

\noindent\textbf{K-CARE (Ours):} Our proposed framework integrates Symmetrical Contextual Anchoring (SCA) with TGKI training and Analogical Prototype Reasoning (APR), extending the model's cognitive reach beyond its parametric memory.

\vspace{-0.5mm}

\subsection{Implementation Detail}
We employed the Qwen3-8B LLM to build K-CARE. All training process on
32 ascend 910B NPUs, with the batch sizes on each NPU as 16. The weight decay is
0.1 and the deepspeed stage is 2. Furthermore, we use the Adamw
optimizer with the learning rate $1e^{-5}$. The max length of the model
training is 690 and the epoch is 1. 
For inference, we use vLLM with the following settings, where temperature is set to 0 for deterministic outputs.

\subsection{Overall Performance}

The evaluation results are summarized in Table~\ref{tab:experiment_overall}. LLM Base performs poorly on three-level tiering, achieving a Macro F1 of 52.94\%, a Weighted F1 of 60.29\%, and an Accuracy of 59.73\%, indicating limited fine-grained relevance discrimination without alignment. More importantly, LLM Base shows an extreme imbalance in irrelevant detection, achieving a very high Bad category precision of 92.42\% but an extremely low Bad category recall of 39.72\%, which reflects the optimistic bias of vanilla LLMs in relevance prediction~\cite{DBLP:conf/www/lref-jd}. After supervised alignment, LLM SFT substantially improves tiering performance, achieving a Macro F1 of 80.97\%, a Weighted F1 of 86.82\%, and an Accuracy of 86.95\%, and yields more balanced irrelevant detection with a Bad category precision of 89.91\%, a Bad category recall of 90.30\%, and a Bad category F1-score of 90.11\%, demonstrating that supervised fine-tuning effectively aligns the model with the relevance criteria. Building on LLM SFT, LLM GRPO further improves by exploiting the potential within the model parameters through on-policy optimization, achieving a Macro F1 of 81.18\%, a Weighted F1 of 87.04\%, and an Accuracy of 87.17\%, together with a Bad category recall of 91.30\% and a Bad category F1-score of 90.48\%, indicating more consistent relevance discrimination and stronger offline identification of bad cases. 

Our proposed K-CARE framework consistently outperforms all baselines across both evaluation dimensions.
Specifically, K-CARE achieves consistent performance gains in three-level classification, reaching a Macro F1 of 82.58\%, a Weighted F1 of 87.99\%, and an Accuracy of 88.05\%, indicating robust multi-level product tiering. 
For Irrelevant Classification, K-CARE achieves a leading Recall of 91.51\% and an F1-score of 91.09\%, bolstering the detection of 'Bad' cases to safeguard user search experience.
Collectively, these gains demonstrate that grounding the model in behavior-derived implicit knowledge and expert-curated prototypes successfully fills the contextual voids within the model's parametric memory, effectively calibrating its decision boundaries for more robust relevance inference.

\subsection{Ablation Study}

Table~\ref{tab:experiment_overall} presents a cumulative ablation of K-CARE. Starting from LLM Base, we progressively add the two main components, SCA and APR. Within SCA, we further conduct a fine-grained cumulative study by incrementally enabling QSAP, PSAQ, and TGKI. Enabling QSAP brings the largest initial improvement, increasing Macro F1 to 82.00\% and Bad F1 to 90.74\%, suggesting that query--product anchoring effectively reduces semantic ambiguity for relevance judgment. Adding PSAQ further improves the overall performance, reaching a Macro F1 of 82.16\% and a Bad F1 of 90.84\%, indicating that symmetric anchoring of query and product benefits fine-grained tiering and bad-case identification. With TGKI, SCA attains a Macro F1 of 82.39\% and a Bad F1 of 90.99\%, showing that task-guided training helps the model better internalize and utilize the injected behavioral context. Finally, adding APR on top of SCA yields the best results, achieving a Macro F1 of 82.58\% and a Bad F1 of 91.09\%, which suggests that analogical reasoning grounded in expert-curated rationales further calibrates decision boundaries on ambiguous cases.

% Table~\ref{tab:experiment_overall} presents a cumulative ablation of K-CARE. Starting from LLM Base, we progressively add the two main components, SCA and APR. Within SCA, we further conduct a fine-grained cumulative study by incrementally enabling QSAP, PSAQ and TGKI.
% The integration of QSAP yields the most significant initial gain (Macro F1 to 82.00\% and Weighted F1 to 87.04\%), confirming that query-product anchoring is fundamental for resolving semantic ambiguity, 
% while the addition of PSAQ further refines performance by enriching product-side intent. With QSAP and PSAQ symmetrically anchor the concept of query and product, K-CARE can achieve a Macro F1 of 82.16\%, a weighted F1 of 87.66\%, an Accuracy of 87.73\% and a Bad F1 of 90.84\%. 
% Introducing TGKI elevates the metrics to a new celling proving that specialized training tasks are essential for the model to effectively internalize  and leverage the provided behavioral context. The whole module of SCA reaches a Macro F1 of 82.39\%, a weighted F1 of 87.85\%, an Accuracy of 87.94\% and a Bad F1 of 90.99\%. 
% Finally, the full model, incorporates SCA and APR, achieves peak performance, validating that analogical inference against authoritative rationales effectively calibrates decision boundaries in ambiguous cases. 

% 北极星 authoritative 这个词又出现啦！！！

\subsection{Online Evaluation}
To meet the strict latency and throughput requirements of production search, we deploy a compact BERT-based model as the online serving model. Rather than directly serving the 8B LLM, we adopt a label distillation strategy where the K-CARE teacher model generates high-quality relevance labels on large-scale unlabeled data, and the student BERT is trained on these labels via standard supervised fine-tuning. This design preserves the knowledge gains from K-CARE while reducing inference latency by an order of magnitude compared to the teacher model, thereby enabling real-time relevance scoring at scale.

We conduct online A/B testing in JD.com’s search engine by randomly splitting a fraction of search traffic into treatment (K-CARE-distilled BERT) and control (production baseline) groups. Two independent control slot (Base1 and Base2) are run to ensure robustness. Each session processes a large volume of search queries, providing sufficient statistical power for detecting meaningful differences. 

To evaluate the online impact of K-CARE, we adopt a controlled query-level comparison protocol. We curate a set of 10,000 representative queries spanning diverse e-commerce categories and route each query through three parallel experiment slots: two control slots (Base1 and Base2) running the production model, and one treatment slot running the K-CARE-distilled BERT. By fixing the query set across slots, this design isolates the model effect from traffic variability, ensuring that any observed difference in experience metrics is attributable to the model change. As our primary online metric, we adopt Bad Case Rate, defined as the proportion of queries whose top-10 ranked results contain at least one Bad-labeled product. This metric directly reflects the user-facing search experience, as irrelevant products appearing in top positions are most detrimental to user experience. As shown in Table \ref{tab:online_badcase}, K-CARE reduces the Bad Case Rate from 12.10\% to 11.3\% , a 5.87\% relative reduction.

\begin{table}[h]
  \caption{Online Bad Case Rate Comparison.}
  \centering
  \label{tab:online_badcase}
  \begin{tabular}{l|c}
    \toprule
    \textbf{Models / Sessions} & \textbf{Bad Case Rate} \\
    \midrule
    Base1 Session & 12.11\% \\
    Base2 Session & 12.09\% \\
    \midrule
    \textbf{Test Session (K-CARE)} & \textbf{11.39\%$^{*}$} \\
    \bottomrule
  \end{tabular}
  \vspace{1pt}
\end{table}

\subsection{Case Study}
We present representative cases to qualitatively illustrate how K-CARE grounds relevance judgments with contextual anchoring and expert-curated reasoning prototypes. Each case highlights one key component of K-CARE: QSAP for query-side disambiguation, PSAQ for product-side intent anchoring, and APR for rule-consistent calibration on ambiguous cases.

\paragraph{Case 1 (QSAP: near-homophone brand disambiguation).}
For the query "\texttt{niutaite} deeply-hydrolyzed formula" (\begin{CJK}{UTF8}{gbsn}纽泰特深度水解配方\end{CJK}), the candidate product is \texttt{niukangte} amino-acid formula (\begin{CJK}{UTF8}{gbsn}纽康特氨基酸配方\end{CJK}), and the ground-truth label is Bad. This is a typical near-homophone ambiguity in Chinese, where the two brand mentions differ by only one Chinese character (\begin{CJK}{UTF8}{gbsn}纽泰特\end{CJK} vs. \begin{CJK}{UTF8}{gbsn}纽康特\end{CJK}), which can easily lead to an overly permissive match. 
QSAP resolves this ambiguity by anchoring the query intent to the top-ranked exposure item, which incorporating implicit behavior-derived knowledge learned by the search system, revealing that \texttt{niutaite} corresponds to the English brand name \texttt{Pepti Junior}.
This evidence confirms that \texttt{niutaite} (Pepti Junior) and \texttt{niukangte} (\texttt{Neocate}) refer to two distinct brands, enabling K-CARE to correctly reject the candidate as Bad.

\paragraph{Case 2 (PSAQ: product intent anchoring against title stuffing).}
For the query "hiking boots", the candidate title is mixed and contains keywords for both hiking boots and wading/creek shoes, but the ground-truth label is Bad. Although these terms co-occur in the title, hiking boots and wading/creek shoes correspond to substantially different usage scenarios and functional requirements. PSAQ introduces click-derived behavioral context that reveals the product's true intent: the clicked queries consistently indicate that the item is primarily a wading/creek shoe rather than a boot-style hiking product. With this product-side intent anchor, K-CARE avoids being misled by title stuffing and correctly rejects the candidate as Bad.

\paragraph{Case 3 (TGKI: enabling effective utilization of behavioral anchors).}
For the query "tf perfume", the candidate product is a \texttt{TIFFANY \& Co.} perfume, and the ground-truth label is Bad. Here, QSAP provides a critical contextual anchor: the top-ranked exposure item is a TOM FORD perfume, indicating that \texttt{tf} refers to TOM FORD rather than TIFFANY\&Co. However, this evidence is only indirectly encoded. The model must infer from the top-ranked product that \texttt{tf} stands for TOM FORD, rather than reading an explicit definition. Without TGKI, the model observes the anchor product but fails to leverage it, instead forcing an incorrect association between \texttt{tf} and TIFFANY, resulting in a false positive prediction of Relevant. With TGKI, the model has been trained to interpret contextual anchors through task-guided objectives, enabling it to recognize that the top-ranked TOM FORD product disambiguates the query intent, and correctly rejects the TIFFANY product as Bad. This case demonstrates that providing enriched context alone is insufficient and task-guided training is essential for the model to effectively utilize indirect behavioral knowledge.

\paragraph{Case 4 (APR: matching rubrics clarify via prototype rules).}
For the query "Feitian Moutai 53\% (JD self-operated)", the candidate product is Moutai Bulao 1994 53\%, and the ground-truth label is Bad. Here, the key difficulty lies in fine-grained series matching within the same brand: the series/type name of a liquor product can be treated as a model-like constraint that requires strict matching. APR retrieves an expert-curated prototype rationale that explicitly encodes this rule and uses it to calibrate the decision boundary, preventing overly optimistic matches between different series (e.g., Feitian vs. Bulao 1994).

Overall, these cases demonstrate that K-CARE can fill the contextual void and calibrate decision boundaries by grounding relevance judgments in behavior-derived anchors and rule-based prototypes, leading to more reliable tiering and fewer top-ranked irrelevant exposures.

%% file: chapter/Conclusion_v2.tex
\section{Conclusion}
In this paper, we identify \textbf{knowledge boundaries} as the primary bottleneck in LLM-based e-commerce search relevance. The absence of domain-specific context within parametric memory leads to persistent failures in complex scenarios.
To bridge this gap, we propose \textbf{K-CARE}, a framework that extends the model's cognitive reach by grounding its reasoning in \textbf{behavior-derived implicit knowledge} via \textbf{Symmetrical Contextual Anchoring (SCA)} and \textbf{expert-curated prototypical knowledge} via \textbf{Analogical Prototype Reasoning (APR)}. 
Extensive offline evaluations on 120,000 real-world samples and online A/B tests demonstrate that K-CARE significantly enhances search quality through precise relevance tiering and robust irrelevant case identification. 
These results validate the effectiveness of integrating external domain intelligence with analogical reasoning to overcome the contextual limitations of general-purpose LLMs in specialized industrial tasks.